# EDUKOI: DEVELOPING AN INTERACTIVE SONIFICATION TOOL FOR ASTRONOMY BETWEEN ENTERTAINMENT AND EDUCATION

*Lucrezia Guiotto Nai Fovino*

University of Padua,
Via Venezia 8,
Padua, Italy
lucrezia.guiottonaifovino@phd.unipd.it

*Massimo Grassi*

University of Padua,
Via Venezia 8,
Padua, Italy
massimo.grassi@unipd.it

*Anita Zanella*

Istituto Nazionale di Astrofisica,
Vicolo dell'Osservatorio 5,
Padua, Italy
anita.zanella@inaf.it

*Michele Ginolfi*

University of Florence,
Via G. Sansone 1,
Sesto Fiorentino (FI), Italy
michele.ginolfi@unifi.it

*Luca Di Mascolo*

University of Trieste,
Via Tiepolo 11,
Trieste, Italy
luca.dimascolo@units.it

## ABSTRACT

Edukoi is a software that aims to make interactive sonification suitable to convey and extract information. The program design is a modification of the software Herakoi, which sonifies images in real time mapping pitch to colour using a motion-aware approach for allowing users to interact with images through sound. The pitch-colour association of Hearkoi, albeit pleasing from the entertainment side, is not efficient for communicating specific information regarding colours and hues to listeners. Hence we modified it to create an instrument to be used by visually impaired and sighted children to explore images through sound and extract accurate information. We aim at building a flexible software that can be used in middle-schools for both art and science teaching. We tested its effectiveness using astronomical images, given the great fascination that astronomy always has on kids of all ages and backgrounds. Astronomy is also considered a very visual science, a characteristic that prevents students from learning this subject and having a related career. With this project we aim to challenge this belief and give to students the possibility to explore astronomical data through sound. Here we discuss our experiment, the choices we made regarding sound mappings, and what psychophysiological aspects we aim to evaluate to validate and improve Edukoi.

## 1. INTRODUCTION

Sonification is the science and the art of translating data into non-speech sound [1]. It can be used to convey information to different audiences, from professionals to children. Sonification can be used in various scientific domains such as, for example, astronomy. Astronomy is typically perceived as a visual science, hence discouraging Blind and Visually Impaired (BVI) people to study it and pursue such a carrier. However there is no reason for this, as astronomical data are collected as numbers that we convert into images to make sense of them. In astronomy, we can have three main applications of sonification: aid astronomers in research; foster the dissemination of science to a non-specialistic public; aid BVI people accessing astronomical knowledge [2]. Most available softwares for image sonification map pitch to colour and are structured for sonifying images following a specific path (ie: reading the image from left to right, or from the center out), allowing for a more complex sound design at the inevitable expense of interactivity, as the images are now "audio-video clips" that have to be enjoyed in the order intended by the sound designers. Here we present Edukoi [3], a tool for interactive sonification that can be used to explore astronomical images. Edukoi was created by modifying the software Herakoi [4] (both are freely available on github). These sonification tools might be useful? for two aims: the dissemination of astronomy to broader audiences and to BVI and aurally-oriented people, and a tool to allow BVI people to explore images, not necessarily astronomy related, in a different way from the audio description.

## 2. INTERACTIVE TOOLS FOR ASTRONOMICAL SONIFICATION

Edukoi is an evolution of Herakoi. Herakoi is a motion-sensing sonification tool that allows to hear the sound of any image, for educational, artistic and entertainment purposes. Our pipeline makes use of the publicity available MediaPipe Hand Landmarker model, that allows the detection of landmarks of the hands in an image, tracking the keypoint localization of 21 hand-knuckle coordinates. According to the MediaPipe documentation, their model was trained on approximately 30K real-world images and on several synthetic hand models imposed over various backgrounds. The model bundle combines the hand landmarks detection model with a palm detection model. The latter locates hands within any input image, thus defining a region of interest within which the hand landmarks detection model identifies specific hand landmarks. The MediaPipe Hand Landmarker task operates on image data that can be either static data or a continuous stream. We use it to track the position of the hand(s) of the user(s) in real-time in the scene observed by a webcam connected to the computer of the user [5]. The model records the coordinates of the hands of the user that are then re-projected onto the pixel coordinates of the chosen image. The visual properties of the "touched" pixels (in the current version, colour, and saturation) are then converted into the sound properties of the chosen instrument, which can be chosen from a virtual MIDI keyboard. [4]. The current mapping used in Herakoi is coherent with the most popular in sonification of astronomical data (i.e., colour – pitch, with red being low pitched and blue being high pitched, brightness – amplitude) is suited to engage the public and offer an entertaining experience, but is not optimal to extract and remember information on colours and hues from the images that the user explores. We decided to create a modified version of this software, which we called Edukoi, to be primarily used in schools for educational purposes. Our idea was to implement a sound mapping that would allow a more intuitive representation of colours, so to make astronomical images possible to "be seen" in the mind. We will now present this educational tool in detail.

### 3. CURRENT EDUKOI IMPLEMENTATION

In building this tool, we wanted to pursue some general objectives:
- let sighted and BVI children understand that sonification makes astronomy accessible, possibly changing their current perspective
- understand if sonification makes the learning of scientific disciplines more engaging for everyone

To ensure that our tool was efficient in sonifying astronomical images, we wanted to assess if users were able to:
- recognize colours
- recognize shapes
- explore and understand the content of real astronomical images

We chose to map the three RGB colours in an associative way. To add memorization and thus recognition even after long periods not using the tool and thus losing familiarisation, we used natural sounds, and mapped red with the sound of a crackling fire, blue with the sound of bubbles and green with birds and rustling leaves. This would allow anyone to easily remember the sound-to-colour coding, by associating timbre to words. This association is also independent of the culture of level of visual impairment of the users (Kim 2021). The RGB coding makes it difficult to analyse spontaneously more complex colours, as the associations are not widely known. We plan to switch the colour-mapping to CMYB with further developing of the program, as this would make it more intuitive for sighted students, who had the chance to study and experiment with subtractive colour theory, and make a more interesting tool for BVI students to explore artistic images, for example in Art classes.

We also chose to spatialise the sounds, to make colour-identification easy when, for example, two colours are played together. When listening with headphones or other stereo facilities, the red is played to the left earpad, the blue to the right one, and the green from both pads.

### 4. TESTING EDUKOI IN A MIDDLE-SCHOOL

We partnered with the middle school "Scuola Media Don Milani", located in Castiglione delle Stiviere (Italy) to test Edukoi. We tested it with 7 classes (about 150 students in total), aged between 11 and fourteen. Our work plan is structured as follows (for more information on the materials, see "Materials" section):
February 2023:
- first round of structured interviews to assess what the ideas of the students about sonification and the accessibility of astronomy are.
- written survey to assess the attitude towards science [6]
- first test using Edukoi in the classroom

Between February and May: the students will use Edukoi freely with their teachers, during science and art classes. Teachers are asked to record the number of lessons in which they will use the tool.

May 2023:
- second test in the classroom
- written survey to assess the attitude towards science, for the second time
- second round of structured interviews

We started the first intervention/test in the classroom with a brief and interactive explanation about what the work of an astronomer is about. We briefly talked about the difference between blue (young) and red (old) stars. We also briefly explained to the students how Edukoi works as well as the sonification mapping. We mentioned the structure of the experiment, namely that we would first use Edukoi on some test images (geometrical shapes) and then on actual astronomical data showing blue and red stars. The screens of the students were covered with a black paper sheet to prevent them from actually seeing the images displayed on the monitor, as we wanted them to only use the sound to recognize colours and shapes.

Each student recorded the answers on a paper questionnaire that we provided. The actual test was about 30 minutes long, structured in three steps as follows.

1. Colour recognition + shape finding:
Three geometrical shapes (circle, square, equilateral triangle) of identical size (in pixel per cm^2) were presented once in green, once in red and once in blue in a nested-randomised order. We provided the students with the list of the shapes they were interacting with and we asked them to recognize the colour (red, green, or blue). The students had to explore the empty space to find the shape they were looking for.

2. Colour recognition + object finding:
Three images of stars (see "Materials" section) were presented twice, one with the original rotation and one rotated by 180 degrees. The students had to find the stars and identify their colour (red or blue). They were asked to write how many blue stars and how many red stars were present in each image. Students were instructed that when they would hear all three colours at the same time, it was index of a very bright colour, so they were in the centre of the star that appeared white in the image (we used the example of light that appears white but when passed through a prism can actually be broken down in all the colours). They needed to explore the more external halo to understand the actual colour of the star. In the case they would hear red and blue at the same time, it was a "magenta" halo, and that had to be counted as a red star.

3. Colour recognition + shape finding and recognition (optional):
Eighteen geometrical shapes with the same area (the same of trial 1) were presented in a nested randomised order for shape and colour. The students had to understand the colour and the shape of the object that was being sonified. This was purposefully long and we did not expect students to finish it.

During the testing phase, we encountered several technical setbacks due to the devices available at school. Sometimes headphones were not working properly, the computers took a long time to load the images, and the sonified area was bigger compared to when we tested it on our personal computers (likely due to the low resolution of the screens available at school), thus more difficult to explore properly. However, the program worked well. Quantitative data, especially from the object finding step, seem impossible to use (due to the technical problems encountered) but the data from steps one and three will give us insight on how well the colour mapping worked and the qualitative data from the preliminary interviews and post-testing feedback from the classes will prove useful in observing pre/post differences in the conceptualization of inclusive science and interest in sonification. This first test was fundamental to understand criticalities and challenges to build a tool that can be functional on a wide range of computers and to better evaluate our next steps. We will be going back for the second phase of testing in May and perform a similar test, after having implemented some solutions for the main technical issues.

## 5. RESULTS AND FEEDBACK

Quantitative results are still being processed. Here we report some qualitative results.

### 5.1. First round of interviews

Most students thought that the best way in which technologies could help BVI people to access scientific knowledge was to cure blindness or to invent special glasses that somehow would allow BVI to see. There was a minority of students suggesting that other senses, such as hearing or touch, could be used. However, most of them did not have any specific idea on how other senses could be used in practice. The vast majority of students seemed very positive about the possibility for BVIs to become scientists, although posing it as an explicit question can definitely give space to some biases. However, more than the answer to this question, it is interesting to see how the knowledge of accessibility issues for BVI people and potential strategies for inclusion will evolve in the students. Even broadening their knowledge about the topic in an unexpected manner such as sonification could be considered a success.

### 5.2. Feedback on sonification

The results about the effectiveness of sonification in conveying the content (e.g., colour, number counts, shapes) of images still have to be processed, but during the experiment students asked us pertinent questions and showed that they understood how to use Edukoi. For example, one of the images [see Figure 3] of the stars showed the halo of the red stars as magenta; thus the sounds played by Edukoi were blue and red at the same time. They noticed it and asked us how to evaluate that. We asked each class to provide some feedback after the test, and it seems they had fun although they found counting the stars very difficult. Noticeably, all classes maintained the overall attention very high during the whole experiment, despite the test being fairly long and repetitive. By analysing the results from the written survey, we will understand if there is any change in the level of engagement with scientific disciplines when using sonification rather than classic educational methods.

The use of the hands made difficult to explore such a large space in a systematic manner, thus it was easy to miss some stars. This tool can probably be more useful to explore bigger images, even with more details, than to find multiple small objects in pictures.

### 5.3. Feedback from a BVI student

a.      We had the chance to collect some feedback from a BVI student and his support teacher. He was enthusiastic about the tasks and very focused, although he found it very difficult to find the sonified areas, as any kind of spatial reference was lacking. To listen to the sound associated with each pixel of an image, the user has to move the hand on a plane parallel to the screen, in front of the camera. Hence it was difficult for the BVI student to understand where the camera and the screen were and to move the hand accordingly. This potential limitation could be overcome by using a physical support (e.g., transparent plexiglass sheet in front of the screen) that the users could touch and use to track the position of their hand with respect to the screen and camera.

## 6. CONCLUSION

The use of interactive sonification tools seems a promising way to make sonification a more widespread concept/tool?. The first testing experience provided useful feedback that will be used to improve the overall design of Edukoi (and, subsequently, Herakoi as an overall project). After the second and final data collection, we will be able to assess if our sonification project changed how students think about the accessibility of science for BVI people and if their overall attitude towards science changed. As a general note, this kind of interactive sonification program may work better paired with visual images or haptic stimuli of some kind.

## 7. ACKNOWLEDGMENT

We would like to thank the students, teachers, and staff from the Scuola Media Don Milani that supported our first testing phase. In particular, we thank the teacher Roberto Veltri and the psychologist Arianna Daldosso for their practical support during the project.

## 8. MATERIALS

### 8.1. Structured interviews

We asked four open questions:
1.      What comes to your mind if I say "science" and "BVI people"?
2.      Do you think that a BVI student has difficulties in studying scientific subjects at school?
3.      How could science and technology help BVI people in studying scientific subjects, such as astronomy?
4.      What would you say to a blind girl/boy of your age that tells you that he/she wants to become an astronomer?
Additional information provided by the students was recorded.

### 8.2. Questionnaire

We used an italian translation, made for this study, of the mATSI-2 questionnaire [6]

### 8.3. Astronomical images

We used two images of stellar fields taken from the archive of the European Southern Observatory [7]:
- the image of the Jewel Box star cluster (NGC 4755) observed with FORS1 on the Very Large Telescope; this image was cropped in two different ways, so that in the first case there were 2 blue stars and 1 red and in the second case there were 3 blue stars and 1 red [8].

- the image of the double star SS Leporis observed with the Very Large Telescope Interferometer PIONEER [9], taken in three different visits.

We blurred the images to decrease their resolution and hence the amount of details (e.g., little stars, spikes) and we increased the saturation to make colours clearer to recognize.

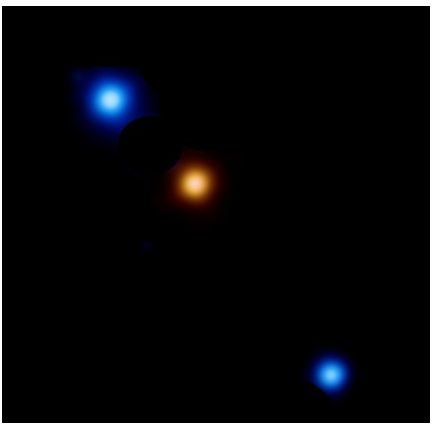

Figure 1a: Jewel Box star cluster (NGC 4755) cropped to have 2 blue stars and 1 red and in the second case there were 3 blue stars and 1 red

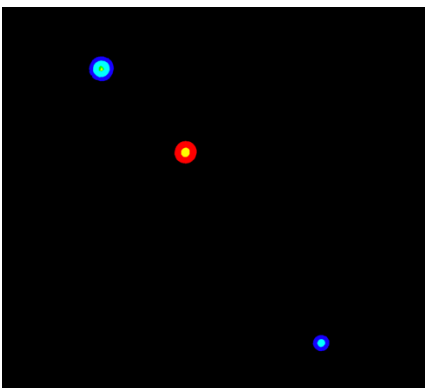

Figure 2b: Figure 1a as appeared in Edukoi

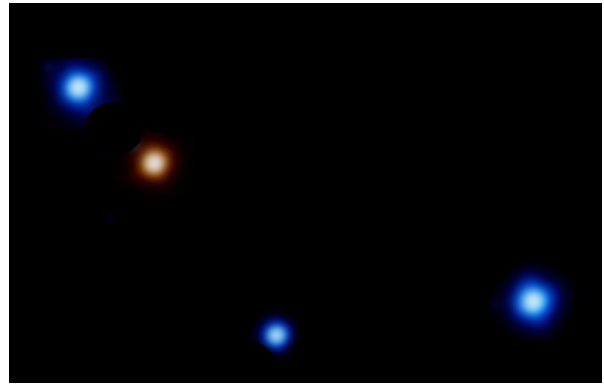

Figure 2: Jewel Box star cluster (NGC 4755) cropped to have 3 blue stars and 1 red

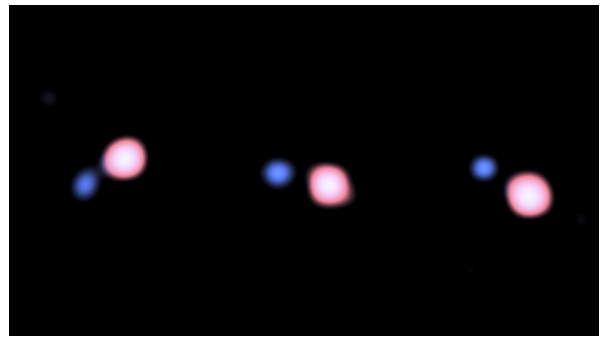

Figure 3: double star SS Leporis